\documentclass[preprint,12pt]{elsarticle} 
\usepackage{graphicx} 
\usepackage{amssymb} 

\journal{Nuclear Physics A} 

\begin{document}

\begin{frontmatter} 

\title{Neutron skins of $^{208}$Pb and $^{48}$Ca from pionic probes} 

\author[a]{E.~Friedman}
\address[a]{Racah Institute of Physics, The Hebrew University, 
Jerusalem 91904, Israel} 

\date{\today}

\begin{abstract}
The neutron skin of $^{208}$Pb has received considerable attention in recent years.
A variety of strongly-interacting probes depict a rather consistent picture 
but pionic probes have not been referred to in this context. We
present here neutron-skin values from pionic atoms and from total reaction 
cross sections of $\pi ^+$ between 0.7 and 2 GeV/c which fit well into the picture.
In addition we show that a neutron skin for $^{48}$Ca can be obtained from
existing data on pionic atoms and the result agrees with pion scattering
experiments and with the scattering of $\alpha $ particles.

\end{abstract}


\end{frontmatter}

\section{Introduction} 
\label{sec:intro}                                              
The topic of neutron skin of nuclei, or the difference between r.m.s. radii
of neutron and proton distributions $\delta R_{np}$, received in recent years
considerable attention because it correlates with several quantities
of interest both for nuclear physics and astrophysics, see Tsang  et al.
\cite{TSC12} for a very recent extensive discussion and references. The
neutron skin of $^{208}$Pb was the subject of several experiments which used
a variety of strong-interaction probes, leading to a rather consistent set
of `accepted' values, see, for example, Jastrz\c{e}bski  et al. \cite{JTL04}. 
A later experiment with
295 MeV polarized protons \cite{ZMS12} agreed,
within errors, with earlier measurements at 800 MeV and with the above
accepted values. In contrast to experiments with strongly-interacting probes
which require a model for the derivation of nuclear sizes,
the PREX experiment at Jlab. aimed at model-independent extraction
of the neutron radius of $^{208}$Pb from parity-violating electron
scattering. Unfortunately, 
it failed, so far, to provide meaningful results due to large statistical errors
\cite{AAA12}.

A strongly-interacting probe that has not been referred to in this context
 is the pion. Among the wealth
of available experimental results on the interaction of pions with Pb we
chose two examples, namely, 
(i) pionic atoms where one can use the strong interaction level shifts 
and widths to gain information on nuclear sizes, 
and (ii) total reaction cross sections of  $\pi ^+$ mesons on nuclei
between momenta of 0.7 and 2~GeV/c. 
It is shown in the present work that these two examples of 
pionic probes lead to values for the neutron
skin of $^{208}$Pb that have acceptable uncertainties and agree well with
the majority of other experiments.

Very recently a new proposal to Jlab. \cite{MMP12} addressed the neutron
skin of $^{48}$Ca for similar reasons as for $^{208}$Pb. Here too pion probes
have not been discussed and therefore we present below also values of $\delta R_{np}$
obtained from analyses of pionic atoms of $^{48}$Ca and compare the
derived values for the neutron skin with other studies with pions. We include also some
remarks on elastic scattering of alpha
particles as a possible source of information on nuclear radii,  arguing that
the results of diffraction scattering
quoted in Ref.\cite{MMP12} are irrelevant to the present problem,
but that, in contrast, experiments extending to large angles that were overlooked
so far, do provide neutron skin values in agreement with pionic atoms.

\section{Neutron skin of $^{208}$Pb}
\label{sec:Pb}

\subsection{Pionic atoms}
\label{sec:piPb}

Strong interaction level shifts and widths of pionic atoms have been measured
and interpreted for the past few decades, see Ref.\cite{FGa07} for a review
and references. An optical potential that is related to nuclear densities is
capable of reproducing the experimental results which, in turn, could supply
information on nuclear densities. Garcia-Recio  et al. \cite{GNO92} 
studied pionic atom potentials while varying also radial parameters of
the neutron distributions. They presented values of $\delta R_{np}$ from their own
analysis as well as from two previous models, leading to a weighted average
of $\delta R_{np}=0.18\pm0.05$ fm for $^{208}$Pb.

\begin{figure}[thb]
\begin{center}
\includegraphics[width=0.8\textwidth]{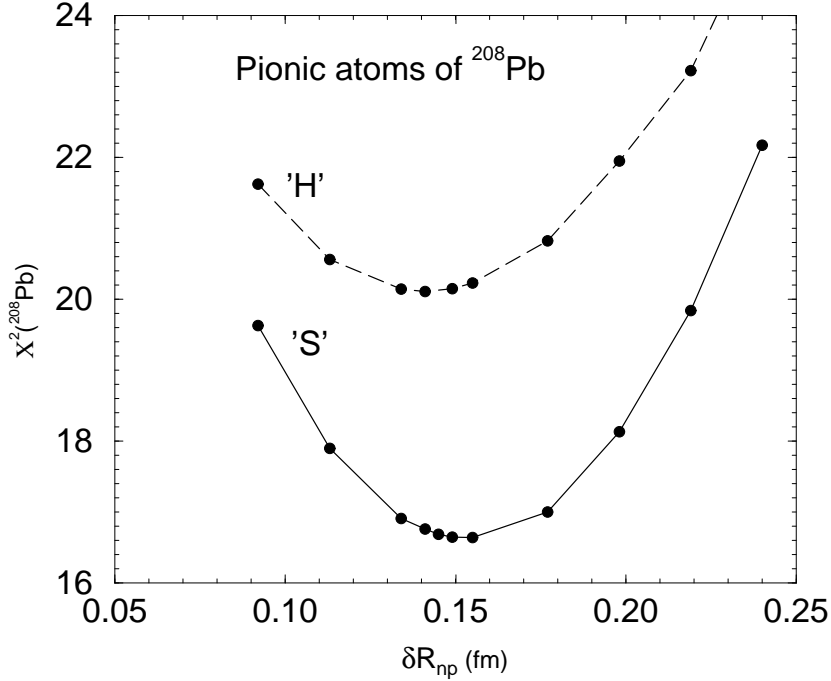}
\caption{$\chi ^2$ values for four data points of
pionic atoms of $^{208}$Pb {\it vs.} $\delta R_{np}$ based on pionic atoms potentials
from global fits \cite{FGa07}. The data points are shifts and widths of the $3d$
and $4f$ levels of pionic atoms of $^{208}$Pb.}
\label{fig:piatPb}
\end{center}
\end{figure}

The present work is based on more recent global fits of potentials
to 100 data points across the periodic table, 
where the  enhancement
of the isovector $s$-wave amplitude $b_1$ was taken care of by introducing
the chiral-motivated density dependence due to Weise \cite{Wei00,Wei00a}.
For more details see Figs. 11 \& 12 of Ref.\cite{FGa07}.
Figure \ref{fig:piatPb} shows values of $\chi ^2$ 
for $^{208}$Pb obtained with potentials
of the global fit, when varying only radial parameters of the neutron distribution.
The experimental results are from de~Laat et al. \cite{LDH91}.
Since the r.m.s. radius does not determine uniquely a distribution we 
used two extreme shapes for the neutron distribution where the excess
r.m.s. radius over the corresponding value for the proton distribution is either
due to an increased half-density radius (S) or increased diffuseness
of the surface (H), referred to as `skin' and `halo', respectively,
by Trzci\'{n}ska et al. \cite{TJL01}. The figure shows
results of fits to four data points when varying one parameter so that the minimal
$\chi ^2$ achieved of 5.5 per degree of freedom is somewhat disappointing compared to
the corresponding value for the global fit
to 100 data points which is only 1.9. This is 
obviously reflected in the uncertainties and the
derived values for $\delta R_{np}$ are 0.15$\pm$0.08 fm for the `S' shape
for the neutron distribution and 0.14$\pm$0.10~fm for the `H' shape.
Both results are in full agreement with other derivations of $\delta R_{np}$ for
$^{208}$Pb from strong-interaction experiments.

\subsection{Total reaction cross sections of $\pi ^+$}
\label{sec:sigmaR}

Interactions of high energy particles with nuclei have been a rich source
of information on nuclear sizes \cite{BFG89}. In the present work we
used total reaction cross sections for $\pi ^+$ mesons on nuclei at six
momenta between 0.7 and 2.0 GeV/c \cite{ABB73}. The method was first to `calibrate'
 the potential with total reaction cross sections
$\sigma _R$ for C and Ca, measured together with Pb in the same experiment, 
where the neutron skin vanishes. Then we included in the analysis the
cross sections for Pb and repeated the process varying the neutron
skin $\delta R_{np}$ value. 

Cross sections were calculated by solving the Klein-Gordon equation \cite{FGa07}
with a `$t\rho$' -type optical potential constructed from the relevant forward
pion-proton c.m. scattering amplitudes $f$:

\begin{equation}
f^{\pm}(0)~=~\frac{k_0}{4\pi}\sigma _T^{\pm}(i+\alpha ^{\pm})
\label{eq:forampl}
\end{equation}
where $k_0$ is the pion-proton c.m. momentum at the relevant energy and
$\sigma _T^{\pm}$ are the pion-proton total cross sections. $\alpha ^{\pm}$
are the ratios of real to imaginary parts of the forward scattering amplitudes.
The scattering amplitudes near the forward direction were written as function
of the momentum-transfer q in the usual way,

\begin{equation}
f^{\pm}(q)~=~f^{\pm}(0)e^{-\beta_{\pm}^2q^2/2}
\label{eq:slope}
\end{equation}
and the slope parameters $\beta_{\pm}$ were then used to fold-in finite-range
Gaussian interaction into the potential in coordinate space \cite{ABB73}.
We compared values of $\sigma _T^{\pm}$ from the 2008 SAID analysis \cite{Sai08}
with the corresponding values of Carter et al. \cite{Car68} used in 
Ref.\cite{ABB73}. Since the differences were very small and in any case one needs
Fermi-averaged values of $\sigma _T^{\pm}$ for constructing the potentials, we used
throughout the present work the Fermi-averaged $\pi^{\pm}$-p total cross sections
of Allardyce et al. \cite{ABB73} together with their values of
$\beta_{\pm}$ and of $\alpha ^{\pm}$. Note that the dependence of calculated
$\pi^{\pm}$-nucleus reaction cross sections on the real potential is extremely small. 
Since `$t \rho$'-type optical potentials usually refer to the pion-nucleus
system, the above c.m. forward amplitudes Eq.(\ref{eq:forampl}) were transformed
into the pion-nucleus system, see e.g. Eq. (4)-(7) of Ref. \cite{FGa07}.

\begin{figure}[thb]
\begin{center}
\includegraphics[width=0.8\textwidth]{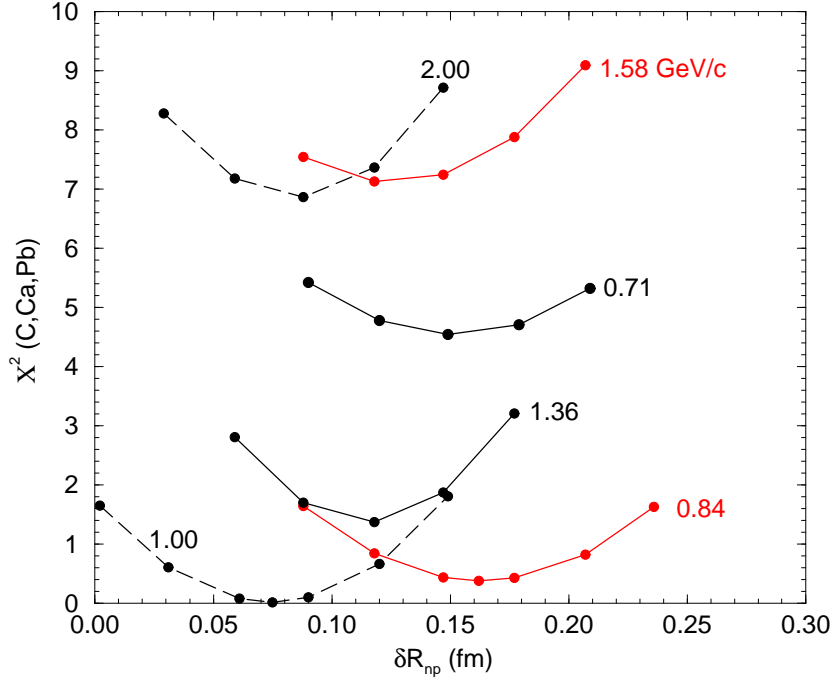}
\caption{$\chi ^2$ values for total reaction cross sections of $\pi ^+$
 on C, Ca and Pb {\it vs.} $\delta R_{np}$
for the six beam momenta indicated. The empirical parameter $B$ is varied at
each momentum, see text.}
\label{fig:pisigmaPb}
\end{center}
\end{figure}

A major point of the present analysis was the estimate of uncertainties.
The first step was to see how well can the above parameter-free potential
reproduce the $\pi ^+$ reaction cross sections on C and Ca, where one may
safely assume that there are essentially no neutron skins. In order to achieve
agreement with experiment we added an empirical imaginary term to the
potential, proportional to the square of the density. Adjusting this single
parameter~(B), $\chi ^2$~values for the two data points were between less than 1.0
and up to 7, depending on momentum. The values of B meant corrections of the order
of $\pm$10\% to the potentials.
Following this calibration we then included, at
each momentum, the $\pi ^+$ reaction cross section for Pb and repeated the
process, varying  radial parameters of the neutron distribution
for Pb. Figure \ref{fig:pisigmaPb} shows results of 
simultanous fits to C, Ca and Pb.
It is reassuring to note that for each momentum the empirical value of B
turned out, at the minimum of $\chi ^2$, to be the same as the corresponding value
found in the calibration. 
Moreover, values of  minima of $\chi ^2$ shown in the 
figure are close
to the corresponding minima  achieved in the calibration 
with the latter obviously affecting
the uncertainties of the derived values of $\delta R_{np}$ for Pb. 
This is simply because the uncertainties are determined not only by the 
curvature of the $\chi ^2$ function but are also scaled by the square root
of the  $\chi ^2$ per degree of freedom, $\chi ^2 /df$, when it is larger
than 1. In this way we included the contributions of the calibration errors 
to the final uncertainties. Table \ref{tab:PbR} summarizes 
the results, and the weighted
average of these is $\delta R_{np}$=0.11$\pm$0.06 fm.
Note that no correlations were found between the empirical parameter B,
the $\chi ^2$ values achieved and the skin values $\delta R_{np}$ obtained
in the present analysis. With the excellent consistency between the calibration
on C and Ca and the eventual application to Pb, the present results seem to
be quite reliable.

\begin{table}
\caption{Values of $\delta R_{np}$ of Pb from $\pi ^+$ reaction cross sections
at different momenta.}
\label{tab:PbR}
\begin{center}
\begin{tabular}{ccc}
p$_\pi$ (GeV/c) & $\chi ^2/df$ & $\delta  R_{np}$ (fm) \\ \hline
0.71 & 2.3 & 0.151$\pm$0.097 \\
0.84 & 0.2 & 0.162$\pm$0.066 \\
1.00 & 0.0 & 0.073$\pm$0.057 \\
1.36 & 0.7 & 0.114$\pm$0.046 \\
1.58 & 3.6 & 0.126$\pm$0.109 \\
2.00 & 3.4 & 0.084$\pm$0.085 \\ \hline
\end{tabular}
\end{center}
\end{table}

The above results were obtained from the experimental cross sections of
\cite{ABB73} for natural isotopic mixture of Pb. At three momenta cross sections
are available also for a target of $^{208}$Pb although with increased
uncertainties. Repeating the analysis for those three cross sections
we obtained values of $\delta R_{np}$ that are consistent with the results
of Table \ref{tab:PbR} but with increased uncertainties.

Finally a comment on the work of Allardyce et al. 
\cite{ABB73} is in order.
The experiment reported in Ref.\cite{ABB73} was intended to obtain information
on neutron densities in nuclei from {\it ratios} of total reaction cross sections
for $\pi ^+$ and $\pi ^-$. The conclusions of that project
of almost vanishing neutron skin in heavy
nuclei had been a puzzle for a long time. An emerging explanation is that the
so-called elastic scattering corrections to the data 
could have had small systematic errors. The corrections are the integrated
elastic scattering cross sections for angles beyond the angle 
subtended by the transmission
counters, to be subtracted from the measured beam-loss cross sections,
before extrapolations to zero solid angle provide the total reaction cross section
\cite{ABB73}. The corrections are usually calculated with an optical potential,
and the program used during the initial phases of the experiment \cite{ASt68}
was found later to have some approximations which were inadequate. For the smallest
angles measured \cite{ABB73} the elastic scattering corrections for $\pi ^+$ are
up to an order of magnitude smaller than the corresponding corrections for $\pi ^-$.
It is therefore believed that the cross sections for $\pi ^+$
used in the present work
are reliable, which is not the case for $\pi ^-$.
Unfortunately the raw data were
 no longer available to
make a re-analysis possible \cite{Bat08}.

\section{Neutron skin of $^{48}$Ca}
\label{sec:Ca}

\begin{figure}[thb]
\begin{center}
\includegraphics[width=0.8\textwidth]{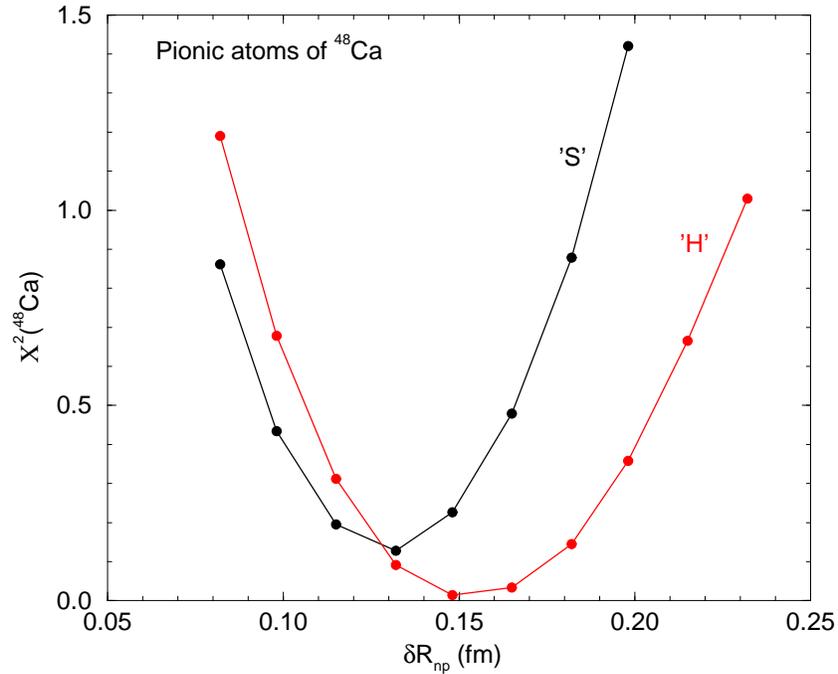}
\caption{$\chi ^2$ values for the shift and width of the $2p$ level of
pionic atoms of $^{48}$Ca {\it vs.} $\delta R_{np}$ based on pionic atoms potentials from
global fits \cite{FGa07}.}
\label{fig:piatCa}
\end{center}
\end{figure}

In the present work we analyzed the $^{48}$Ca pionic atoms data 
of Powers et al. \cite{PWH80} in a similar way to the analysis presented
above for pionic atoms of $^{208}$Pb using the potentials of the latest global
analysis \cite{FGa07}. Figure \ref{fig:piatCa} shows $\chi ^2$ values for 
the strong interaction 2$p$ level shift and width varying only the radial extent
of the neutron distribution. As for $^{208}$Pb we used both the `S' and the `H'
versions of the neutron density \cite{TJL01} and 
the best-fit values for the neutron skin are $\delta R_{np}$=0.13$\pm$0.06~fm
for the `S' version and $\delta R_{np}$=0.16$\pm$0.07~fm for the `H' version.
Note that pionic atoms of $^{48}$Ca have not been included in the 
 analysis of Ref.\cite{GNO92}.

An extensive study of r.m.s. radii of Ca isotopes with pionic 
probes was carried out
by Gibbs and Dedonder \cite{GDe92} who analyzed scattering of $\pi ^+$ and $\pi ^-$
across the (3,3) resonance. Our pionic atoms results are in very good agreement
with their value of $\delta R_{np}$=0.11$\pm$0.04~fm for $^{48}$Ca. 
Earlier application of pions to 
study radii of Ca isotopes yielded similar result \cite{JbC77}.

\section{Discussion}
\label{sec:Dis}

Pionic probes have been shown to provide information on neutron skins in nuclei,
$\delta R_{np}$, which is a topic of current interest. For $^{208}$Pb we have
obtained values for $\delta R_{np}$ from the latest global analysis of pionic atoms
across the periodic table and from analysis of total reaction cross section of $\pi ^+$
in the 0.7 to 2 GeV/c momentum range. The results are in agreement with each other
and with the majority of other results from strongly-interacting probes.

The neutron skin of $^{48}$Ca is receiving renewed attention 
in connection with the C-REX proposal to Jlab. \cite{MMP12}
which is analogous to the PREX experiment \cite{AAA12}.
The present work demonstrates that
the pionic atoms method yields values for the neutron skin which agree with results
from pion scattering by $^{48}$Ca across the (3,3) resonance.
It is worth mentioning that radial parameters of the Ca isotopes have been studied
in great detail by scattering of 104 MeV $\alpha $ particles under the 
necessary conditions for obtaining such information, namely, scattering to large
angles well beyond the diffraction region. As an example we quote 
one result for $^{48}$Ca \cite{GRF84} with 
$\delta R_{np}$=0.17$\pm$0.05~fm, in very
good agreement with the present work. It will therefore be possible to compile
strong-interaction values for $\delta R_{np}$ also for $^{48}$Ca
 in anticipation of a forthcoming C-REX experiment \cite{MMP12}.

\section*{Acknowledgements} 

Stimulating discussions with A. Gal and correspondence with C.J. Batty are gratefully
acknowledged.

\end{document}